# Exploring the Pattern Formation of Lysozyme Drying Droplets in Phosphate Buffer Saline Solution

*Anusuya Pal[1], Amalesh Gope[2*], and Germano S. Iannacchione[1]*

[1]*Order-Disorder Phenomena Laboratory, Department of Physics, Worcester Polytechnic Institute, Worcester, MA, 01609, USA*
[2]*Department of English, Tezpur University, Tezpur, Assam, 784028, India*

*Email: amaleshtezu@gmail.com*

**Abstract**

The process of drying is a simple physical mechanism that drives a system to relax from one equilibrium point to another. The native states of the constituent particles in the droplets can be linked to the emergent morphological patterns via this drying process. This paper explores the interplay between different initial concentrations of a globular protein lysozyme ($\emptyset_p$) and the salts ($\emptyset_s$) in the phosphate buffer saline (PBS). The $\emptyset_s$ = 0 wt% embodies the lysozyme solution prepared in de-ionized water. The samples at $\emptyset_s$ = 0.9 wt% display a dark texture in the central region. We examined the drying evolution and dried morphology by extracting the mean gray values (*I*) and standard deviation (*SD*). For this, $\emptyset_p$ was fixed at 9.0 wt%, and only the $\emptyset_s$ were varied. The *I* decreases, and the *SD* increases as the salt crystals start appearing during the drying process. The phase separation of these salts directly maps with this textural evolution and is influenced by the salts' amount in these droplets. The scanning electron microscopic images of the dried films at $\emptyset_s$ = 0.5 wt% at different $\emptyset_p$ show that the lysozyme-salt interactions drop off in the ring region. This ring becomes more apparent with the increasing $\emptyset_p$. Though all the droplets show "coffee-ring" behavior, the leftover lysozyme particles indicated with the mound-like structure, diminishes in the salts' presence. The alteration of the crack patterns is also found. Therefore, the chemistry between multiple salts and lysozyme at various initial concentrations reveals that the features are not merely a sum (or average) of individual microscopic actions. It appears to involve both protein-protein and protein-salt interactions partially averaged over one length/time scale that sets the next larger/longer length/time scale in such drying droplets.

## 1. Introduction

Understanding the emergence of the patterns in the biosystems is hard to predict as well as far complicated than any other soft systems. It is because many of the biological activities are interconnected and dependent on vast parameter space. The mesoscopic or the macroscopic resulting patterns are mostly sensitive to the initial conditions, preparation of the samples, perturbing fields, etc. [1]. Therefore, it is often recommended to systematically study these systems by changing one parameter at a time. The patterns in these bio-systems emerge due to the local self-assembling interactions between the constituent particles. One such example is when the system is relaxed from one state (initial colloidal fluid) to another (dried organized film) through a non-equilibrium (drying) process that exchanges energy and matter with the environment to drive this kind of assembly.

Prior efforts in the drying droplet community show how the different amounts of salts affect the pattern formation in two globular protein samples, lysozyme and bovine serum albumin (BSA) [2-6]. Gorr et al. [2] have studied the lysozyme protein at various concentrations of NaCl. This study reported the presence of three distinct regions in the company of NaCl. The first one is formed in the peripheral ring, where most of the lysozyme is present. The second one forms different salts structures that occupy the secondary ring area (observed adjacent to the ring), and the final one is observed in the central regions. A much like observation is also reported in BSA-saline protein drying droplets by Yakhno [3]. This study infers that the salt crystals are phase-separated by forming different zones from homogeneous protein film near the periphery to the salt crystals in the central region. Recently, Pathak et al. [4] investigated the effects of multiple salts ($MgCl_2$ and KCl) on the BSA patterns. This study reveals that the crystal structures depend on the initial tuning ratio of these salts. New and more sophisticated image processing techniques have become widespread because of the importance of images in such drying studies. The pattern recognition tools, such as k-means clustering and k-nearest neighbor algorithm, were applied by Gorr et al. [5]. It shows that these tools are powerful enough to differentiate the lysozyme-NaCl deposits based on the salts' initial concentration. Carreón et al. [6] have used first- and second-order statistics to specify textural image properties and explored information about the evolution of the final state of drying BSA-lysozyme films in NaCl salts' presence.

Despite the intense research on protein-saline drying droplets, to our best knowledge, no systematic study is being performed to understand multiple salts' effects on various concentrated lysozyme protein solutions. This current paper investigates the drying droplet consisting of lysozyme in different phosphate buffer saline solution using bright-field and scanning electron microscopy. A morphological grid is displayed by changing the initial concentration of lysozyme ($\emptyset_p$) along the x-axis and varying the initial concentration of the salts ($\emptyset_s$) present in the buffered saline along the y-axis. To better understand these morphological patterns, the microstructural analysis is done by varying $\emptyset_p$ and fixing $\emptyset_s$. Furthermore, the drying evolution of different initial concentrations of lysozyme at a fixed $\emptyset_s$ is investigated and quantified using textural image processing techniques. Therefore, this paper attempts to answer a few fundamental questions; do we always observe three regions in lysozyme-saline droplets? If not, why? What are the effects of multiple salts on lysozyme droplets? Does it behave similarly as reported in the BSA droplet? Can the texture analysis reveal information about the protein-protein and protein-salt interactions? However, this study does not intend to provide an exhaustive understanding of the field's state on pattern formation in these biological (protein) systems. Instead, it aims to motivate the budging researchers to consider bio-colloidal drying systems relating to macroscopic and microscopic entities when an initial fluid consumes its internal chemical potential as the solvent evaporates and drives these complex microscopic structures' self-assembly.

## 2. Materials and Experimental Methods

The hen-egg white lysozyme (HEWL) is a globular protein used extensively by researchers for *in-vitro* studies. It is because the characteristics of HEWL mimics with the lysozyme found in human mucosal secretions, such as tears, saliva, etc. The lyophilized form of HEWL (Catalog number L6876) was purchased from Sigma Aldrich, USA. The globular compact nature of HEWL is maintained as most of the hydrophobic residues are buried inside the core. In contrast, numerous positively and negatively charged residues are exposed at the lysozyme's surface [7].

The different concentrations of the phosphate buffer saline (PBS) were prepared by diluting 1x to 0.75, 0.5, and 0.25x. The 1x PBS solution contains 0.137M (~8.0 mg/mL) NaCl, 0.002M (~0.2 mg/mL) KCl, and 0.0119M (~1.44 mg/mL of $Na_2HPO_4$ and ~0.24 mg/mL of $KH_2PO_4$) phosphates at a pH of ~7.4. It is purchased from Fisher BioReagents, USA (Catalog number BP24384). The various amounts, i.e., 100, 75, 50, 35, 25, and 10 mg of lysozyme, are weighed and mixed separately in 1 mL of these PBS solutions. Therefore, the samples were prepared with the initial concentrations of HEWL ($\emptyset_p$ = 9.0, 6.9, 4.8, 3.3 2.4, and 1.0 wt%) at the initial concentrations of the various salts ($\emptyset_s$ = 0.9, 0.7, 0.5, 0.2, and 0.0 wt%) present in the phosphate buffer saline (PBS). The $\emptyset_s$ = 0 wt% means that the lysozyme solution is prepared without dissolving it in any PBS, but in the de-ionized water (Millipore, 18.2 MΩ.cm at 25 °C). This indicates that the direct comparison of the prior studies' results [2-6] is not possible with our study.

A volume of ~1 μL sample solution is pipetted on a freshly cleaned coverslip (Catalog number 48366-045, VWR, USA) under ambient conditions (the room temperature of ~25 °C and the relative humidity of ~50%). The images of the dried samples were captured within 24 hours. The drying evolution is monitored at every two seconds only for those samples, where the initial lysozyme concentration is kept fixed ($\emptyset_p$ = 9 wt%) and the initial salts' concentrations ($\emptyset_s$) are varied from 0.9 to 0.0 wt%. The drying evolution of the contact angle of these samples was also measured using a contact angle goniometer (Model number 90, Ramé-hart Instrument Company, Succasunna, NJ, USA). The clock started when the droplets were deposited on the coverslips. The images were captured under 5x magnification using bright-field optical microscopy (Leitz Wetzlar, Germany) configured in the transmission mode. An 8-bit digital camera (Model number MU300, Amscope) is attached to the microscope to click the top-view images. All these experiments were repeated three times, and these samples show the highest reproducibility. The textural evolution of these droplets during the drying process is done by extracting the mean gray values (*I*) and the standard deviation (*SD*) using ImageJ [8]. The *oval tool* of ImageJ is selected to capture the area of interest in such droplets. The scanning electron microscopy (JEOL-7000F, JEOL Inc., MA, USA) is used for the samples by varying $\emptyset_p$ of 9.0, 4.8, 3.3, and 1.0 wt% at a fixed $\emptyset_s$ of 0.5 wt% and ($\emptyset_p, \emptyset_s$) = (9.0, 0.0 wt%). The ~4 nm layer of gold nanoparticles was sputter-coated, and the images were captured at the accelerating voltage of 3 kV and the probe current of 5 mA.

3. Results

Figure 1 displays a morphological grid of the samples varying the initial concentrations of lysozyme ($\emptyset_p$) from 9.0 to 1.0 wt% and the initial concentrations of various salts present in the PBS ($\emptyset_s$) from 0.9 to 0.2 wt%. The $\emptyset_s$ = 0 wt% embodies the lysozyme solution prepared in the de-ionized water. Though all these deposits show the "coffee-ring" effect [10], the diverse patterns are observed for each $\emptyset_p$ and $\emptyset_s$. The lysozyme films show a mound-like structure when the solution is prepared without any external salts. A dimple (or depression) is also noticed within this mound. The mound area gets wider as the $\emptyset_p$ increases. The random cracks are only observed in the peripheral ring at ($\emptyset_p, \emptyset_s$) = (1.0, 0.0) wt%. However, these cracks are spread throughout the film as the $\emptyset_p$ increases. The radial and orthoradial cracks promote well-connected (small and large) domains in these droplets. Some fringes appear in the concentrated lysozyme samples at $\emptyset_s$ = 0 wt%. Many domains in the ring get delaminated, which are predominantly observed at

($\emptyset_s$, $\emptyset_p$) = (0.0, 9.0) wt%. The concentration dependence of these lysozyme droplets in the salts' absence is detailed in our previous publication [7]. A comparison of these patterns reveals that the mound diminishes in the salts' presence. However, no general trend in these patterns is observed at a fixed $\emptyset_s$. Interestingly, such a trend is noticed when $\emptyset_p$ is fixed and $\emptyset_s$ is varied. For example, at $\emptyset_p$ = 1.0 wt%, the ring width decreases with the increasing $\emptyset_s$. The central region becomes grainy. The texture becomes darker, and some thread-like structures appear in the central region. The samples at $\emptyset_s$ = 0.9 wt% display a dark texture in the central region and a gray texture in the peripheral ring. The various textural regions are observed in the central region. The multiple rings are also found in the highly concentrated lysozyme samples. The random small cracks are observed at $\emptyset_p$ = 1.0 wt% whereas, mostly radial cracks are found in the peripheral ring as $\emptyset_p$ increases in the salts' presence. In contrast, no overall drift is found whether these cracks intervene in the central region. For instance, the sample at ($\emptyset_p$, $\emptyset_s$) = (9.0, 0.9 wt%) shows crack patterns, whereas other samples at $\emptyset_s$= 0.9 wt% do not. In the presence of the low salt concentrations, different cracks are observed in the lysozyme droplets in the central region at $\emptyset_p \geq 6.9$ wt%. Though these optical images showcase these patterns globally, it is to be noted that these do not reveal any microstructural information.

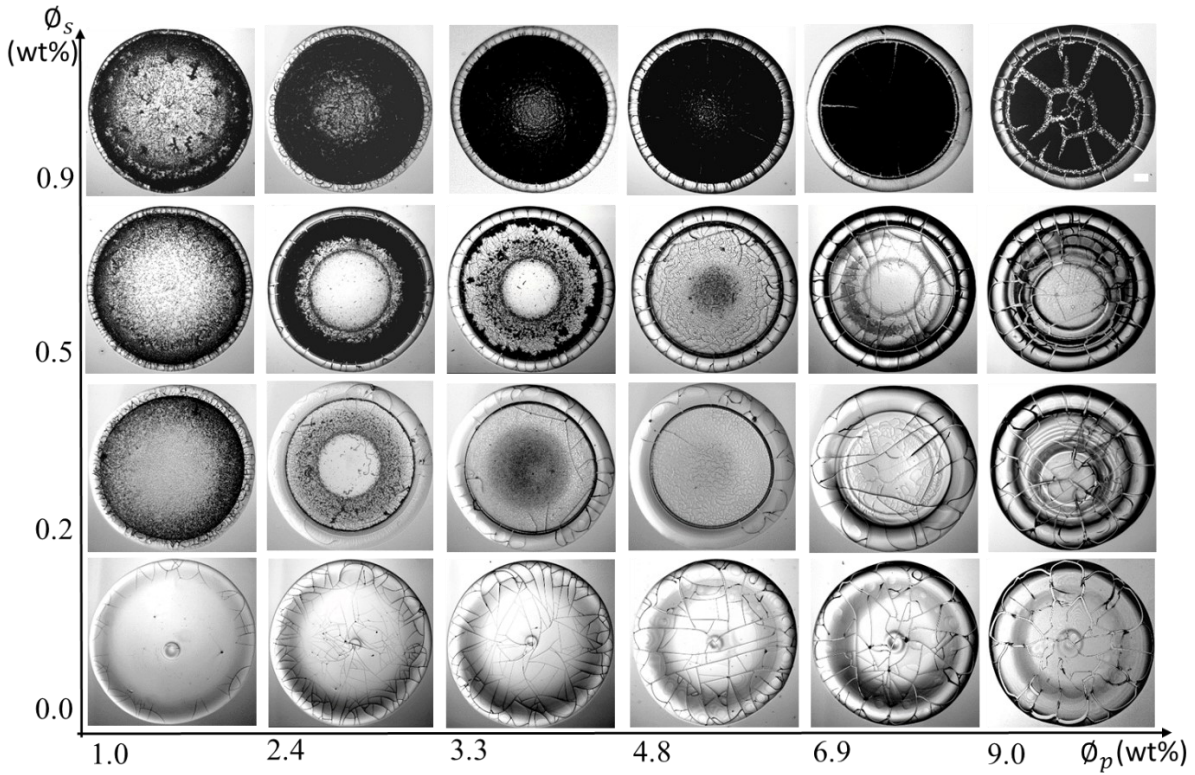

**FIG 1.** Morphological patterns in the drying droplets of lysozyme are displayed. These patterns are shown for the various initial concentrations of lysozyme ($\emptyset_p$) at the initial concentrations of salts present in the phosphate buffer saline ($\emptyset_s$). The $\emptyset_s$ = 0 wt% means that it is prepared in the de-ionized water. The scale bar is of length 0.15 mm.

Figure 2(I-V) exhibits the microstructures of the various concentrated lysozyme samples at $\emptyset_s$ = 0.5 wt%. The sample at ($\emptyset_p$, $\emptyset_s$) = (9.0, 0.0) wt% is displayed in Fig. 2(V). The different regions in the central and peripheral regions were emphasized in all these samples. The sample shows a uniform homogeneous texture in the absence of any external salts [see Fig. 2(V)]. In contrast, the distinct texture is observed at different

regions in the salts' presence [see Fig. 2(IV)]. Not only that, but some non-uniform structures are also uncovered in the crack lines separating the periphery and the central regions. Comparing the central and the peripheral regions, the smooth texture is mostly found in the peripheral ring. However, some snowflakes-like structures appear in the inner ring of the periphery at $(\emptyset_p, \emptyset_s)$ = (3.3, 0.5) wt% [see Fig. 2(II)]. The crystal-like structures are discovered in the zone between the central and peripheral regions at $(\emptyset_p, \emptyset_s)$ = (1.0, 0.5) wt% [see Fig. 2(I)]. These structures are not so prominent as we move towards the central region of the film. The central region is mostly replaced with different forms of the dendrite structures; long but thin structures at $(\emptyset_p, \emptyset_s)$ = (3.3, 0.5) wt% whereas shorter but thicker structures at $(\emptyset_p, \emptyset_s)$ = (9.0, 0.5) wt% in [see Figs. 2(II) and 2(IV)]. The middle region (between peripheral and the central regions) is mostly occupied with a grainy amorphous layer in these samples. On the other hand, it is hard to differentiate this layer between the middle and the central regions at $(\emptyset_p, \emptyset_s)$ = (4.8, 0.5) wt% [see Fig. 2(III)].

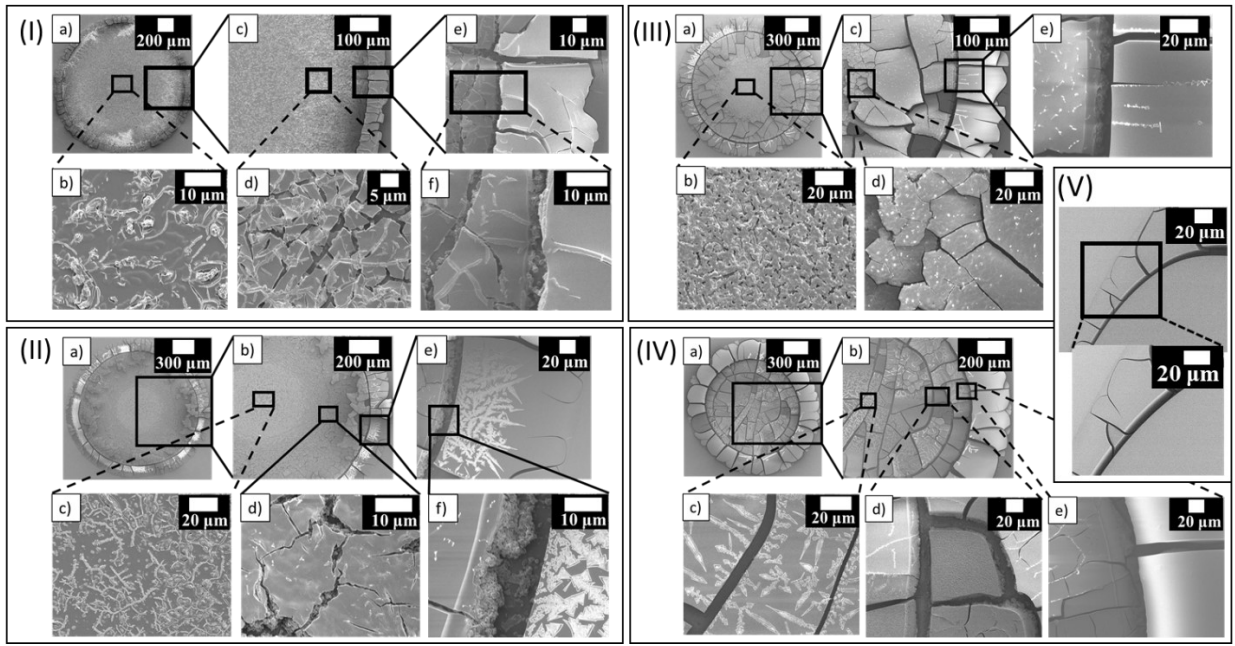

**FIG 2.** Microscopic images of the dried lysozyme samples are displayed. $(\emptyset_p, \emptyset_s)$ = (1.0, 0.5) in (I), $(\emptyset_p, \emptyset_s)$ = (3.3, 0.5) in (II), $(\emptyset_p, \emptyset_s)$ = (4.8, 0.5) in (III), $(\emptyset_p, \emptyset_s)$ = (9.0, 0.5) in (IV), and $(\emptyset_p, \emptyset_s)$ = (9.0, 0.0) wt% in (V). The different length scales are shown as the scale bars in the upper-right corner of each image.

To understand how these distinct structures appear in different regions, we examined the drying evolution and dried morphology by keeping $\emptyset_p$ at 9.0 wt%, and only the $\emptyset_s$ is varied from 0.2 to 0.9 wt%. Figure 3(A-D) describes the drying evolution of the lysozyme droplets, where the first set of images were captured for the solution prepared in the de-ionized water ($\emptyset_s$ = 0 wt%). A uniform gray texture with a dark peripheral band is observed in all the droplets when the first image is captured of the drying process [see Fig. 3(A)]. As time progresses, the fluid front moves from the periphery to the central region [see Fig. 3(B)]. Surprisingly, the texture of the front movement changes in the salts' presence. Once the peripheral ring emerges, the grainy texture starts developing in the central region. Clearly, a distinction is visible at the interface of the inner peripheral ring. At $\emptyset_s$ = 0.2 wt%, the development of the dark texture is not predominantly observed, whereas the darkness increases as the $\emptyset_s$ rises. Simultaneously, the cracks propagate from the periphery towards the center. However, the propagation is not smooth, unlike $\emptyset_s = 0$ wt% and found to be interrupted in the presence of the salts. The mound-like structure begins in the last stage of this fluid front movement, but the salts' presence diminishes its formation in the central region. The multiple rings are found as the front's radius gets smaller [see Fig. 3(B-C)]. The final morphological patterns after the visible drying process are captured in Fig. 3(D).

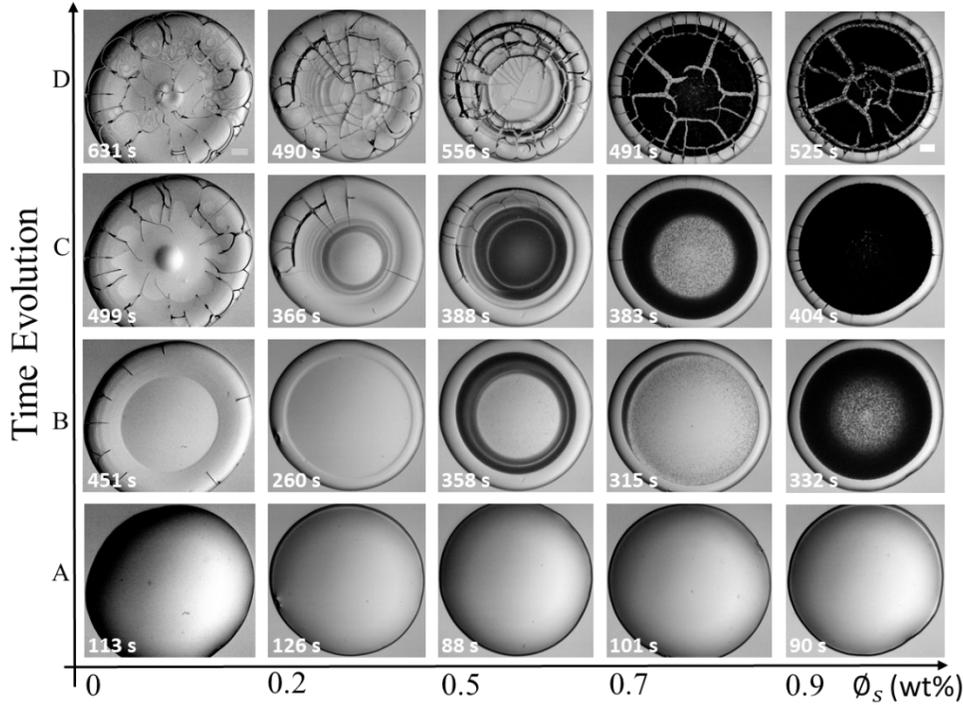

**FIG 3.** The time evolution of lysozyme droplets ($\emptyset_p$ = 9 wt%) at various initial concentrations of salts in PBS ($\emptyset_s$) during the drying process is displayed in A-D. The $\emptyset_s = 0$ wt% embodies the lysozyme solution prepared in de-ionized water. The timestamps are shown at the left-bottom of each image. The white rectangle represents a scale bar of length 0.15 mm in the top-right.

Figure 4(I-IV) shows the quantitative analysis of the textural evolution during the drying process. The first-order statistical parameters, the mean gray values (*I*), and the standard deviation (*SD*) are displayed as a function of the drying time (in seconds) at $\emptyset_s$ ranging from 0.2 to 0.9 wt%. It is to be noted that these parameters describe the gray level distribution of the image's pixel intensity. The *I* defines the averaged values, whereas the *SD* illustrates the textural complexity. The *I* stays nearly constant at the beginning of the drying process. It reduces and then fluctuates [marked with a star in Fig. 4(I-IV)]. In contrast, the *SD* starts decreasing linearly till ~300 s and rapidly rises till ~375 s. It decreases, grows again [marked with a star in Fig. 4(I-IV)]. Finally, both the *I* and *SD* saturate in the later phase of the drying evolution. Interestingly, both the *I* and *SD* exhibit significant changes for ~150 s, i.e., between ~300 and ~450 s. The *I* varies within ~15 a.u. at $\emptyset_s = 0.2$ wt%. This change in *I* increases with the rise of $\emptyset_s$; for instance, the *I* reduces from ~50 to ~20 a.u at $\emptyset_s = 0.9$ wt%. The images [shown in Fig. 4(I-IV)] show that the variation in these textural parameters (*I* and *SD*) occurs considerably when the dark textured fluid front moves in the central region.

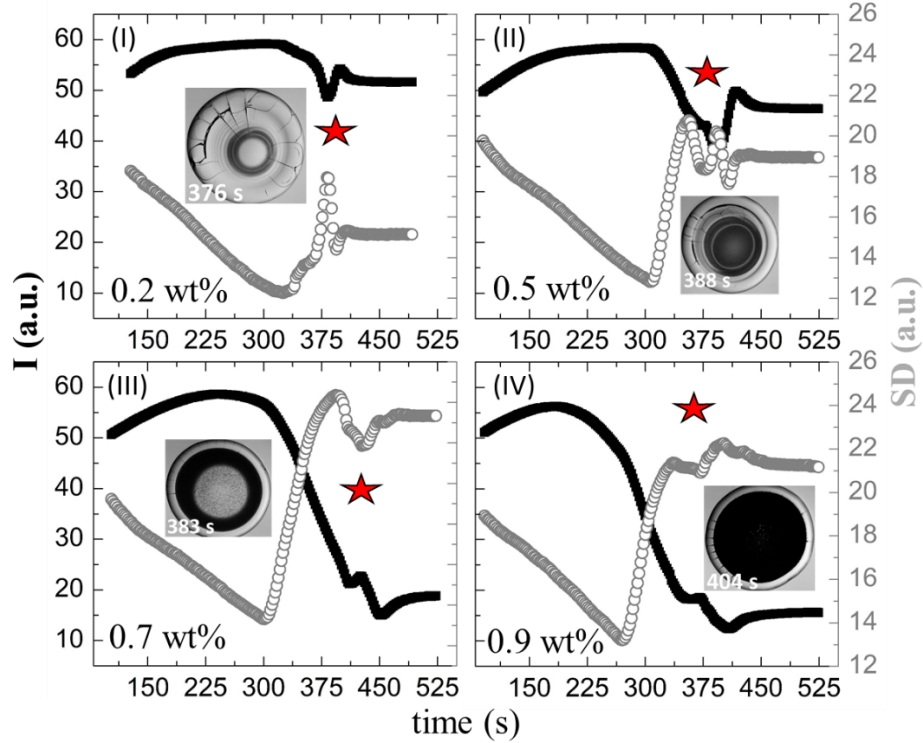

**FIG 4.** Textural analysis of the lysozyme droplets ($\emptyset_p = 9$ wt%) at various initial concentrations of salts in PBS ($\emptyset_s$) during the drying process is displayed in (I-IV). The x-axis defines the drying time. The left-y (shown in back color squares) and right-y axes (shown in gray color circles) in each graph describe the mean gray values (*I*) and the standard deviation (*SD*) in arbitrary units (a.u.), respectively. The star symbol indicates the parameters' fluctuations. The optical image represents the droplets' morphology when significant changes in these parameters are observed.

## 4. Discussions

As soon as the droplets are pipetted on the coverslip, the height and the contact angle start reducing (checked with the contact angle goniometer). The non-uniform textural gradient in the optical images [Fig. 3(A)] and the images captured using the goniometer show that these droplets are partially wetted (the contact angle is ~50 degrees within ~45 s of their deposition) and are of the spherical-cap shape. The curvature of these droplets induces the highest mass-loss near the periphery compared to the central region. The droplets get pinned to the coverslip, and the lysozyme particles are transported through the outward capillary radial flow to compensate for this loss. The process leads to a well-known "coffee-ring effect" [10] like other bio-colloids [1-6]. As time progresses, it is observed that the fluid front recedes from the periphery to the central region, and simultaneously, the contact angle reduces, unlike the result reported in [2]. The deposits in the crack lines (Fig. 2) suggests that there exists a discontinuity at the ring interface [also evident in Fig. 3(B-C)]. It could be speculated that a significant amount of water gets evaporated by this time, and the salts start crystallizing. Since the images were taken in the transmission mode, the thick film gives rise to the dark texture. The dark textured front starts engulfing the central region (a similar phenomenon of the phase transition reported in [3]. The significant fluctuations in the textural evolution [Fig. 4(I-IV)] are also observed at this phase. The complexity (*SD*) increases as the salt crystals (inhomogeneities) begin to appear. As these droplets are pinned, the mechanical stress is relieved by virtue of crack propagation. The appearance of the salt crystals in different lysozyme concentrations affects crack formation [Fig. 1]. It also alters the interaction between the lysozyme particles and changes their aggregation and precipitation processes (samples with and without adding external salts, evident from Figs. 1-2). It also demonstrates that the three prominent regions may or may not be found in the lysozyme droplets in the salts' presence (unlike reported in [2]). Furthermore, the exclusive variation of the salt content might not provide us a clear picture. The chemistry between multiple salts and lysozyme at various initial concentrations is the crucial factor. It is not merely a sum (or average) of individual microscopic actions [Fig. 2]. It appears to involve both protein-protein and protein-salt interactions partially averaged over one length/time scale that sets the next larger/longer length/time scale in such drying droplets.

## 5. Conclusion

This paper reveals that the tuning between lysozyme and salts is essential in determining the morphological patterns. The textual evolution indicates that the interactions between different lysozyme particles during the drying process are dependent on the amount of the salts present in phosphate buffer saline. It also shows that the occurrence of three distinct regions is not the general characteristic of lysozyme-saline droplets; rather, it depends on the relative initial concentrations of lysozyme and salts. This paper rasters the parameter space by varying their initial concentrations and portraying a resulting morphological grid. These protein model systems would benefit us to understand the complexity in other bio-fluids. Therefore, the current research work needs to be continued by creating a corpus of image-databank of such resulting macroscopic patterns. It demands an in-depth exploration to address the fundamentals of "why" and "how," establishing a general understanding of this emerging field of research.

## 6. Acknowledgments

The authors express their gratitude to the *Tinkerbox* program, sponsored by WIN (*Women Impact Network*) at WPI, for the financial supports. We also would like to thank the Department of Physics, WPI.